\documentstyle[osa,manuscript]{revtex}
\begin{document}
\title{Theoretical status of the $B_{c}$ meson in the shifted $l$-expansion
technique }
\author{Sameer M. Ikhdair\thanks{%
sameer@neu.edu.tr}}
\address{$^{\ast }$Department of Electrical and Electronic Engineering, Near East\\
University, Nicosia, North Cyprus, Mersin-10, Turkey}
\maketitle

\begin{abstract}
In the framework of phenomenological static and QCD-motivated model
potentials for heavy quarkonium, we compute the $\overline{b}c$ mass
spectrum as well as its ${\rm 1S}$ hyperfine splitting using the recently
introduced shifted $l$- expansion technique. We also predict the leptonic
constant \ $f_{B_{c}}$ of the lightest pseudoscalar $B_{c}$ and $%
f_{B_{c}^{\ast }}$ of the vector $B_{c}^{\ast }$ states taking into account
the one-loop and two-loop QCD corrections. Further, we use the scaling
relation to predict the leptonic constant of the ${\rm nS}$-states of the $%
\overline{b}c$ system. For the sake of comparison, we use the same fitting
parameters of our previous model potentials. From our results we conclude
that shifted $l$-expansion method has the same accuracy, convergence and
status as our previous work.


\end{abstract}

% INITIALIZE - DONT CHANGE % %  %

\section{ INTRODUCTION}

\noindent The spectrum and properties of the $\overline{b}c$ quarkonium
system have been calculated various times in the past in the framework of
heavy quarkonium theory [1]. Moreover, the recent discovery [2] of $B_{c}$
meson (the lowest pseudoscalar $^{1}S_{0}$ state of the $B_{c}$ system)
opens up new theoretical interest in the subject [1,3-8]. The Collider
Detector at Fermilab ${\rm (CDF)}$ Collaboration quotes $M_{B_{c}}=6.40_{\pm
0.13}^{\pm 0.39}$ ${\rm GeV}$ [2]. This state should be one of a number of
states lying below the threshold for emission of ${\rm B}$ and ${\rm D}$
mesons. Further, such states are very stable in comparison with their
counterparts in charmonium ($\overline{c}c)$ and bottomonium ($\overline{b}%
b) $ systems. A particularly interesting quantity should be the hyperfine
splitting that as for $\overline{c}c$ case seems to be sensitive to
relativistic and subleading corrections in the strong coupling constant $%
\alpha _{s}$. For the above reasons it seems worthwhile to give a detailed
account of the Schr\"{o}dinger energies for $\overline{c}c$, $\overline{b}b$
and $\overline{b}c$ meson systems below the continuum threshold.

Because of the success of the nonrelativistic potential model and the
flavour independence of the $\overline{q}_{1}q_{2}$ potential, we choose a
wide class of phenomenological and a QCD-motivated potentials by insisting
upon strict flavor-independence of its parameters. We also use a potential
model that includes running coupling constant effects in both the
spherically symmetric potential and the spin-dependent potentials to give a
simultaneous account of the properties of the $\overline{c}c$, $\overline{b}%
b $ and $\overline{b}c$ heavy quarkonium systems. Since one would expect the
average values of \ the momentum transfer in the various quark-antiquark
states to be different, some variation in the values of the strong coupling
constant and the normalization scale in the spin-dependent potential should
be expected. To minimize the role of flavor-dependence, we use consistent
values for the coupling constant and universal QCD renormalization scale for
each of the levels in a given system.

Recently, Kwong and Rosner [7] predicted the masses of the lowest vector and
pseudoscalar states of the $\overline{b}c$ system using an empirical mass
formula and a logarithmic potential. Eichten and Quigg [1] gave a more
comprehensive account of the energies and decays of the $B_{c}$ system that
was based on the QCD-motivated potential of Buchm\"{u}ller and Tye [9].
Gershtein {\it et al. }[8] also computed the energies and decays of the $%
B_{c}$ system using the QCD sum-rule calculations. Baldicchi and Prosperi
[6] have fitted the entire light-heavy quarkonium spectrum and computed the $%
\overline{b}c$ spectrum based on an effective mass operator with full
relativistic kinematics$.$ Fulcher [4] extended the treatment of the
spin-dependent potentials to the full radiative one-loop level using the
renormalization scheme developed by Gupta and Radford [10]. Ebert {\it et
al. }[1] investigated the $B_{c}$ meson masses and decays in the
relativistic quark model. Very recently, we reproduced the $\overline{c}c,$ $%
\overline{b}b$ and $\overline{b}c$ spectroscopy by applying the shifted
large-${\rm N}$ expansion technique (SLNET) on the nonrelativistic and
relativistic wave equations using a group of static and improved QCD
motivated potentials [11].

Encouraged by the success of SLNET application on heavy quarkonium systems
[11,12], we extend the previous works in [11] by applying the shifted $l$-
expansion technique (SLET) [13] on the Schr\"{o}dinger equation to reproduce
the $\overline{b}c$ spectroscopy. The SLET has been recently introduced to
solve mathematically the two and three-dimensional Schr\"{o}dinger equation
for spherically and cylindrically symmetric potentials [13]. Further, the KG
and Dirac equations with radially symmetric Lorentz vector and Lorentz
scalar potentials have also been solved [13]. We anticipate that, by working
SLET, we would be able to obtain a better understanding on the status,
convergence and accuracy of this method among the other methods [11]. We
also consider it as a complementary investigation to our previous works
[11]. Further, our results will enable us to check clearly that SLET is just
a parallel perturbative expansion method, i.e., $\overline{l}$-expansion
procedure in a similar manner to $\overline{k}$-expansion procedure.

The contents of this article are as follow: in section \ref{SWE}, we present
the solution of the Schr\"{o}dinger equation using the SLET for the the
non-self conjugate $\overline{Q}q$ meson mass spectrum. In section \ref{SMP}
we briefly present the potentials used. In section \ref{LC} we give the
first-loop and second-loop correction of the $B_{c}$ leptonic decay
constant. Finally, a discussion and conclusion appear in section \ref{DAC}.

\section{ SCHR\"{O}DINGER WAVE EQUATION}

\noindent \label{SWE}In previous papers [11,12] we have applied the shifted
1/N expansion technique (SLNET) to solve nonrelativistic and relativistic
wave equations. The method starts by writting the original wave equation in
an N-dimensional space which is sufficiently large and using expansion $1/%
\overline{k}$ as a perturbation parameter [14]. Here $\overline{k}=N+2l-a,$ $%
N$ being the number of spatial dimensions of interest, $l$ the angular
quantum number, and $a$ is a suitable shift as an additional degree of
freedom and is responsible for speeding up the convergence of the resulting
energy series. In this work another method called the shifted $l$-expansion
technique (SLET) which is simply consists of using $1/\overline{l}$ as an
expansion parameter, where $\overline{l}=l-a,$ $l$ is an angular quantum
number, $a$ is a suitable shift which is mainly introduced to avoid the
trivial case $l=0.$ The choice of $a$ is physically motivated so that the
next to the leading energy eigenvalue series vanish as in SLNET. The method
does not require writting the $N$-dimensional form of the wave equation and
we expand it directly through the quantum number $l$ involved in the
problem. This method seems more flexible and simpler in treatment and has a
quite different mathematical expansion than SLNET. Like SLNET, SLET is also
a pseudoperturbative technique. The radial part of the Schr\"{o}dinger
equation for an arbitrary spherically symmetric potential $V(r)$ (in units $%
\hbar =1)$ 
\begin{equation}
\left\{ -\frac{1}{4\mu }\frac{d^{2}}{dr^{2}}+\frac{l\left( l+1\right) }{4\mu
r^{2}}+V(r)\right\} u(r)=E_{n,l}u(r),  \label{1}
\end{equation}
where $\mu =\left( m_{q}m_{Q}\right) /(m_{q}+m_{Q})$ is the reduced mass for
the two constituent interacting particles and $E_{n,\ell }$ denotes the
Schr\"{o}dinger binding energy. Equation (1) can be rewritten as

\begin{equation}
\left\{ -\frac{1}{4\mu }\frac{d^{2}}{dr^{2}}+\frac{\left[ \overline{l}%
^{2}+\left( 2a+1\right) \overline{l}+a\left( a+1\right) \right] }{4\mu r^{2}}%
+V(r)\right\} u(r)=E_{n,l}u(r),  \label{2}
\end{equation}
where $\overline{l}=l-a$ with $a$ representing a proper shift to be
calculated later. It is clear from Eq. (2) and our previous work [11] [cf.
Eq. (1) in hep-ph/0303182] that $\overline{l}$ is as much as $\overline{k}/2$
of Imbo {\it et al}. [14]. We follow the shifted $l$-expansion method [13]
by defining

\begin{equation}
V(y(r_{0}))\;=\stackrel{\infty }{%
%TCIMACRO{\underset{m=0}{\sum }}%
%BeginExpansion
\mathrel{\mathop{\sum }\limits_{m=0}}%
%EndExpansion
}\left( \frac{d^{m}V(r_{0})}{dr_{0}^{m}}\right) \frac{\left( r_{0}y\right)
^{m}}{m!Q}\overline{l}^{-(m-4)/2},  \label{3}
\end{equation}
and also the energy eigenvalue expansion [11,12]

\begin{equation}
E_{n,l}\;=\stackrel{\infty }{%
%TCIMACRO{\underset{m=0}{\sum }}%
%BeginExpansion
\mathrel{\mathop{\sum }\limits_{m=0}}%
%EndExpansion
}\frac{\overline{l}^{(2-m)}}{Q}E_{m}.  \label{4}
\end{equation}
Here $y=\overline{l}^{1/2}(r/r_{0}-1)$ with $r_{0}$ is an arbitrary point
where the Taylor expansions is being performed about and $Q$ is a scale to
be set equal to $\overline{l}^{2}$ at the end of these calculations.
Inserting Eqs. (3) and (4) into Eq. (2) yields

\[
\left[ -\frac{1}{4\mu }\frac{d^{2}}{dy^{2}}+\frac{1}{4\mu }\left( \overline{l%
}+\left( 2a+1\right) +\frac{a(a+1)}{\overline{l}\;}\right) \stackrel{\infty 
}{%
%TCIMACRO{\underset{m=0}{\sum }}%
%BeginExpansion
\mathrel{\mathop{\sum }\limits_{m=0}}%
%EndExpansion
}\frac{(-1)^{m}(m+1)y^{m}}{\overline{l}^{m/2}}\right. 
\]

\begin{equation}
+\left. \frac{r_{0}^{2}}{Q}\stackrel{\infty }{%
%TCIMACRO{\underset{m=0}{\sum }}%
%BeginExpansion
\mathrel{\mathop{\sum }\limits_{m=0}}%
%EndExpansion
}\left( \frac{d^{m}V(r_{0})}{dr_{0}^{m}}\right) \frac{\left( r_{0}y\right)
^{m}}{m!}\overline{l}^{(2-m)/2}\right] \chi _{n_{r}}(y)=\xi _{n_{r}}\chi
_{n_{r}}(y).  \label{5}
\end{equation}
Hence the final analytic expression for the $1/\overline{l}$ expansion of
the energy eigenvalues appropriate to the Schr\"{o}dinger particle is [13] 
\begin{equation}
\xi _{n_{r}}=\frac{r_{0}^{2}}{Q}\stackrel{\infty }{%
%TCIMACRO{\underset{m=0}{\sum }}%
%BeginExpansion
\mathrel{\mathop{\sum }\limits_{m=0}}%
%EndExpansion
}\overline{l}^{(1-m)}E_{m}.  \label{6}
\end{equation}
Now we formulate the SLET for the nonrelativistic motion of spinless
particle bound in spherically symmetric potential $V(r)$. Equation (6) is
exactly of the type of Schr\"{o}dinger like equation for the one dimensional
anharmonic-oscillator and has been investigated for spherically symmetric
potential by Imbo {\it et al.} [14]:

\begin{eqnarray}
\xi _{n_{r}} &=&\overline{l}\left[ \frac{1}{4\mu }+\frac{r_{0}^{2}V(r_{0})}{Q%
}\right] +\left[ (n_{r}+\frac{1}{2})\;\omega +\frac{(2a+1)}{4\mu }\right] 
\nonumber \\
&&+\frac{1}{\overline{l}}\left[ \frac{a(a+1)}{4\mu }+\gamma ^{(1)}\right] +%
\frac{\gamma ^{(2)}}{\overline{l}^{2}}+O\left( \frac{1}{\overline{l}^{3}}%
\right) ,  \label{7}
\end{eqnarray}
where the expressions $\gamma ^{(1)}$and $\gamma ^{(2)}$are given explicitly
in Appendix A. Thus, comparing Eq. (6) with Eq. (7) gives 
\begin{equation}
E_{0}=V(r_{0})+\frac{Q}{4\mu r_{0}^{2}},  \label{8}
\end{equation}
\begin{equation}
E_{1}=\frac{Q}{r_{0}^{2}}\left[ \left( n_{r}+\frac{1}{2}\right) \omega +%
\frac{(2a+1)}{4\mu }\right] ,  \label{9}
\end{equation}
\begin{equation}
E_{2}=\frac{Q}{r_{0}^{2}}\left[ \frac{a(a+1)}{4\mu }+\gamma ^{(1)}\right] ,
\label{10}
\end{equation}
and 
\begin{equation}
E_{3}=\frac{Q}{r_{0}^{2}}\gamma ^{(2)}.  \label{11}
\end{equation}
The quantity $r_{0}$ is chosen so as to minimize the leading term, $E_{0}$,
that is, [12]

\begin{equation}
\frac{dE_{0}}{dr_{0}}=0\text{ \ \ \ }{\rm and}\text{ \ \ }\frac{d^{2}E_{0}}{%
dr_{0}^{2}}>0,  \label{12}
\end{equation}
which yields the relation

\begin{equation}
Q=2\mu r_{0}^{3}V^{\prime }(r_{0}).  \label{13}
\end{equation}
Further, to solve for the shifting parameter $a$, the next contribution to
the energy eigenvalues is chosen to vanish [11-14], i.e., $E_{1}=0,$ which
provides smaller contributions for the higher-order corrections in (4)
compared to the leading term contribution (8). It implies that the energy
states are being calculated by considering only the leading term $E_{0\text{ 
}},$ the second-order, $E_{2\text{ }}$and the third-order, $E_{3\text{ }}$%
corrections. Therefore, the shifting parameter is determined via 
\begin{equation}
a=-\frac{\left[ 1+2\mu (2n_{r}+1)\omega \right] }{2},  \label{14}
\end{equation}
with

\begin{equation}
\omega =\frac{1}{2\mu }\left[ 3+\frac{r_{0}V^{\prime \prime }(r_{0})}{%
V^{\prime }(r_{0})}\right] ^{1/2}.  \label{15}
\end{equation}
Thus, the Schr\"{o}dinger binding energy (4) to the third order is

\begin{equation}
E_{n,l}=V(r_{0})+\frac{1}{r_{0}^{2}}\left[ \frac{a(a+1)+Q}{4\mu }+\gamma
^{(1)}+\frac{\gamma ^{(2)}}{\overline{l}}+O\left( \frac{1}{\overline{l}^{2}}%
\right) \right] .  \label{16}
\end{equation}
Further, setting $\overline{l}=\sqrt{Q}$ which rescales the potential, we
derive an analytic expression that satisfies $r_{0}$ as

\begin{equation}
2l+\left[ 1+(2n_{r}+1)\left[ 3+\frac{r_{0}V^{\prime \prime }(r_{0})}{%
V^{\prime }(r_{0})}\right] ^{1/2}\right] =2\left[ 2\mu r_{0}^{3}V^{\prime
}(r_{0})\right] ^{1/2},  \label{17}
\end{equation}
where $n_{r}=n-1$ is the radial quantum number. Once $r_{0}$ is being found
through Eq. (17) for any arbitrary state, the determination of the binding
energy for the $\overline{Q}q$ system becomes relatively simple and
straightforward. Finally, the Schr\"{o}dinger binding mass can be determined
via

\begin{equation}
M(\overline{Q}q)=m_{q}+m_{Q}+2E_{n,l}.  \label{18}
\end{equation}
It is being found that for a fixed $n,$ the computed energies become more
accurate as $l$ increases [11-14]. This is expected since the expansion
parameter $1/\overline{l}$ becomes smaller as $l$ becomes larger since the
parameter $\overline{l}$ is proportional to $n$ and appears in the
denominator in higher-order correction.

\section{ SOME MODEL POTENTIALS}

\label{SMP}

The $\overline{b}c$ system that we investigate is often considered as
nonrelativistic system [1] and consequently our treatment is based upon
Schr\"{o}dinger equation with a Hamiltonian [1,4] 
\begin{equation}
H_{o}=-\frac{\nabla ^{2}}{4\mu }+V({\bbox r})+V_{{\rm SD}},  \label{19}
\end{equation}
where we have supplemented our nonrelativistic Hamiltonian with the standard
spin-dependent terms [1,11,15] 
\begin{equation}
\text{ \ }V_{{\rm SD}}\longrightarrow V_{{\rm SS}}=\frac{32\pi \alpha _{s}}{%
9m_{q}m_{Q}}({\bbox s}_{1}.{\bbox s}_{2})\delta ^{3}({\bbox r}).  \label{20}
\end{equation}
Here, the spin dependent potential is simply a spin-spin part [1,15] that
would enable us to make some preliminary calculations of the energies of the
lowest two ${\rm S}$-states of the $\overline{b}c$ system. The potential
parameters in this section are all strictly flavor-independent and fitted to
the low-lying energy levels of $\overline{c}c$ and $\overline{b}b$ systems.
Like most authors (cf. [1]), we determine the coupling constant $\alpha
_{s}(m_{c}^{2})$\footnote{%
Due to the lack of any experimental splitting data on the $B_{c}$ meson, as
there is one established state, we have fitted the coupling constant to
reproduce the available $c\overline{c}$ splittings.} from the well measured
hyperfine splitting for the ${\rm 1S}(\overline{c}c)$ state [16]

\begin{equation}
\Delta E_{{\rm HF}}({\rm 1S},\exp )=M_{J/\psi }-M_{\eta _{c}}=117.2\pm 1.5~%
{\rm MeV},  \label{21}
\end{equation}
and for the ${\rm 2S}(\overline{c}c)$ state [16-18] 
\begin{equation}
\Delta E_{{\rm HF}}({\rm 2S},\exp )=M_{\psi ^{\prime }}-M_{\eta _{c^{\prime
}}}=32\pm 14~{\rm MeV},  \label{22}
\end{equation}
for each desired potential to produce the center-of-gravity (cog) of the $%
\overline{M}_{\psi }({\rm 1S})$ value. The numerical value of $\alpha _{s}$
is found to be dependent on the potential form and also be compatible with
the other measurements [1,3,4,6-8]. \ Therefore, the ${\rm 1S}$-state
hyperfine splitting [1,11,15] is given by\footnote{%
To the moment, the only measured splitting of ${\rm nS}-$levels is that of $%
\eta _{c}$ and $J/\psi ,$ which allows us to evaluate the so-called SAD
using $\overline{M}_{\psi }({\rm 1S})=(3M_{J/\psi }+M\eta _{c})/4$ and also $%
\overline{M}({\rm nS})=M_{V}({\rm nS})-(M_{J/\psi }-M\eta _{c})/4n$ [15,19]$%
. $}

\begin{equation}
\Delta E_{{\rm HF}}=\frac{8\alpha _{s}}{9m_{c}m_{b}}\left| R_{{\rm 1S}%
}(0)\right| ^{2},  \label{23}
\end{equation}
with the radial wave function at the origin is determined via [15]

\begin{equation}
\left| R_{{\rm 1S}}(0)\right| ^{2}=2\mu \left\langle \frac{dV(r)}{dr}%
\right\rangle .  \label{24}
\end{equation}
Hence, the total mass of the low-lying pseudoscalar $B_{c}$ meson is [11]

\begin{equation}
M_{B_{c}}(0^{-})=m_{c}+m_{b}+2E_{1,0}-3\Delta E_{{\rm HF}}/4,  \label{25}
\end{equation}
and for the vector $B_{c}^{\ast }$ meson

\begin{equation}
M_{B_{c}^{\ast }}(1^{-})=m_{c}+m_{b}+2E_{1,0}+\Delta E_{{\rm HF}}/4.
\label{26}
\end{equation}
Hence, the square-mass difference can be simply found as

\begin{equation}
\Delta M^{2}=M_{B_{c}^{\ast }}^{2}(1^{-})-M_{B_{c}}^{2}(0^{-})=2\Delta E_{%
{\rm HF}}\left[ m_{c}+m_{b}+2E_{1,0}-\Delta E_{{\rm HF}}/4\right] .
\label{27}
\end{equation}
The perturbative part of such a quantity was evaluated at the lowest order
in $\alpha _{s}.$ Baldicchi and Prosperi [6] used the standard running QCD
coupling expression

\begin{equation}
\alpha _{s}({\bbox Q})=\frac{4\pi }{\left( 11-\frac{2}{3}n_{f}\right) \ln
\left( \frac{{\bbox Q}^{2}}{\Lambda ^{2}}\right) }.  \label{28}
\end{equation}
with $n_{f}=4$ and $\Lambda =0.2~{\rm GeV}$ cut at a maximum value $\alpha
_{s}(0)=0.35,$ to give the right $J/\psi -\eta _{c}$ splitting (21) and to
treat properly the infrared region [6]. Further, Brambilla and Vairo [3]
took in their perturbative analysis $0.26\leq \alpha _{s}(\mu =2$ ${\rm GeV}%
)\leq 0.30.$ After the observation of the $\eta _{c}({\rm 2S})$ meson [17],
Badalian and Bakker [18] determination of the coupling constant $\alpha _{%
{\rm HF}}(\mu _{1},{\rm 1S})\simeq 0.335$ is rather large with $\mu
_{1}\simeq \frac{1}{2}M(J/\psi )=1.55${\rm \ }${\rm GeV}$ and $\alpha _{{\rm %
HF}}(\mu _{2},{\rm 2S})\simeq 0.18$ with $\mu _{2}\simeq 2M(\psi ^{\prime
})=7.4${\rm \ }${\rm GeV}$ which implies the very large value for the
renormalization scale. They [20] also used $\alpha _{s}(\mu =0.92$ ${\rm GeV}%
)\simeq 0.36$ for all states, but the splittings do practically not change
if $\alpha _{s}(\mu =1.48$ ${\rm GeV})=0.30$ is taken. This result shows
that the strong coupling constant depends on the renormalization scale and
it changes from one state to another [18]. Further, Motyka and Zalewski [21]
found $\alpha _{s}(m_{c}^{2})=0.3376$ and from which they calculated $\alpha
_{s}(m_{b}^{2})=0.2064$ and $\alpha _{s}(4\mu _{\overline{b}c}^{2})=0.2742.$
Therefore, it is quite clear that the coupling constant is dependent on the
quarkonium system.

The commonly used potentials are of two types: (i) pure phenomenological and
(ii) partly phenomenological, but motivated by both perturbative and
non-perturbative QCD at short and long distances. In this work we use the
following types of potentials:

\subsection{Static potentials}

It is seen that the most phenomenological potentials in Eq. (19) may be
gathered up in general form [21,22]: 
\begin{equation}
V(r)=-ar^{-\alpha }+br^{\beta }+c\text{ \ \ \ \ }0\leq \alpha ,\beta \leq 1,%
\text{ \ }a,b\geq 0.  \label{29}
\end{equation}
where $c$ may be of either sign. The mixed powerlaw (29) comprises more than
ten potentials given by Refs. [21,22]. It seems quite reasonable that the
short range behaviour of the quarkonium potential is less singular than $%
-r^{-1}$ and the confining potential does not rise quickly as $r$ due to the
screening effects of quark pair creation. Therefore, the effective
quarkonium potential consist of two terms, one of which, $%
V_{v}(r)=-ar^{-\alpha },$ transforms like a time-component of a Lorentz
4-vector and the other, $V_{s}(r)=br^{\beta }+c,$ like a Lorentz scalar. We
limit our study to the $\alpha =\beta =\nu $ case

\begin{equation}
V(r)=-ar^{-\nu }+br^{\nu }+c\text{ \ \ \ \ }0\leq \nu \leq 1,\text{ \ }%
a,b\geq 0.  \label{30}
\end{equation}
proposed by Lichtenberg et al. [23]. Like most authors [1,4,11,23], we
consider a class of static potential which give reasonable accounts for the $%
\overline{c}c$ and $\overline{b}b$ spectra. This comprises a wide class of
potentials presented explicitly in our previous works [11] some are
QCD-motivated potentials like Cornell ($\nu =1),$ and other typical pure
phenomenological potentials like Song-Lin ($\nu =1/2),$ Turin ($\nu =3/4$),
Logarithmic ($\nu \rightarrow 0)$ which are all belonging to the class (30)
and Martin ($\alpha =0,\beta =0.1)$ belonging to the class (29). The
motivation of this choice is that all of these potentials lie very close
together in the range of distances $0.1\leq r\leq 1,$ which is the
characteristic interval of $\overline{c}c$ and $\overline{b}b$ spectra.

\subsection{QCD-motivated potentials\ \ \ \ \ \ \ \ \ \ \ \ \ \ \ \ \ \ \ \
\ \ \ \ \ \ \ \ \ \ \ \ \ \ \ \ \ \ \ \ \ \ \ \ \ \ \ \ \ \ \ \ \ \ \ \ \ \
\ \ \ \ \ }

\subsubsection{\it Igi-Ono potential}

Buchm\"{u}ller and Tye [9] proposed a potential which is consisting of two
parts, at short distances the two-loop perturbative calculation of the
interquark one-gloun exchange [24]: 
\begin{equation}
V_{{\rm OGE}}^{(n_{f}=4)}(r)=-\frac{16\pi }{25}\frac{1}{rf(r)}\left[ 1-\frac{%
462}{625}\frac{lnf(r)}{f(r)}+\frac{2\gamma _{E}+\frac{53}{75}}{f(r)}\right] ,
\label{31}
\end{equation}
with

\begin{equation}
f(r)=\ln \left[ \frac{1}{r^{2}\Lambda _{\overline{MS}}^{2}}+b\right] ,
\label{32}
\end{equation}
where $n_{f}=4$ is the number of flavors with mass below $\mu $ and $\gamma
_{E}=0.5772$ is the Euler's number. Moreover, at long distances the
interquark potential grows linearly leading to confinement as 
\begin{equation}
V_{L}(r)=ar.  \label{33}
\end{equation}
Therefore, the Igi-Ono potential is [24] 
\begin{equation}
V^{(n_{f}=4)}(r)=V_{{\rm OGE}}^{(n_{f}=4)}+ar+dre^{-gr}~,  \label{34}
\end{equation}
where the term $dre^{-gr}$ in (34) is added to interpolate smoothly between
the two parts and to adjust the intermediate range behavior by which the
range of $\Lambda _{\overline{MS}}$ is extended keeping linearly rising
confining potential. Numerical calculations show that potential is good for $%
\Lambda _{\overline{MS}}$ in the range $100$-$500~{\rm MeV}$ keeping a good
fit to the $\overline{c}c$ and $\overline{b}b$ spectra. The QCD coupling
constant $\alpha _{s}$ in (20) is defined in the Gupta-Radford (GR)
renormalization scheme [10]

\begin{equation}
\alpha _{s}=\frac{6\pi }{\left( 33-2n_{f}\right) \ln \left( \frac{\mu }{%
\Lambda _{{\rm GR}}}\right) },  \label{35}
\end{equation}
where the scale parameter in the Gupta-Radford (GR) renormalization scheme $%
\Lambda _{GR}$ [10] is related to $\Lambda _{\overline{MS}}$ by 
\begin{equation}
\Lambda _{{\rm GR}}=\Lambda _{\overline{MS}}\exp \left[ \frac{49-10n_{f}/3}{%
2\left( 33-2n_{f}\right) }\right] .  \label{36}
\end{equation}
Thereby, the three types of this potential are displayed in [11].

\subsubsection{\it Improved Chen-Kuang potential}

Chen and Kuang [25] proposed two improved potential models so that the
parameters therein all vary explicitly with $\Lambda _{\overline{MS}}$ so
that these parameters can only be given numerically for several values of $%
\Lambda _{\overline{MS}}.$ Such potentials have the natural QCD
interpretation and explicit $\Lambda _{\overline{MS}}$ dependence both for
giving clear link between QCD and experiment and for convenience in
practical calculation for a given value of $\Lambda _{\overline{MS}}$. It
has the general form 
\begin{equation}
V^{(n_{f}=4)}(r)=kr-\frac{16\pi }{25}\frac{1}{rf(r)}\left[ 1-\frac{462}{625}%
\frac{lnf(r)}{f(r)}+\frac{2\gamma _{E}+\frac{53}{75}}{f(r)}\right] ,
\label{37}
\end{equation}
where the string tension is related to Regge slope by $k=\frac{1}{2\pi
\alpha 
%TCIMACRO{\UNICODE[m]{0xb4}}%
%BeginExpansion
{\acute{}}%
%EndExpansion
}$. The function $f(r)$ in (37) is 
\begin{equation}
f(r)=ln\left[ \frac{1}{\Lambda _{\overline{MS}}r}+5.10-A(r)\right] ^{2},
\label{38}
\end{equation}
with

\begin{equation}
A(r)=\left[ 1-\frac{1}{4}\frac{\Lambda _{\overline{MS}}}{\Lambda _{\overline{%
MS}}^{I}}\right] \frac{1-\exp \left\{ -\left[ 15\left[ 3\frac{\Lambda _{%
\overline{MS}}^{I}}{\Lambda _{\overline{MS}}}-1\right] \Lambda _{\overline{MS%
}}r\right] ^{2}\right\} }{\Lambda _{\overline{MS}}r}.  \label{39}
\end{equation}
The scale parameter $\Lambda _{\overline{MS}}^{I}$ is very close to the
value of $\Lambda _{\overline{MS}}$ determined from the two-photon processes
and is also close to the world-averaged value of $\Lambda _{\overline{MS}}.$
The fitted values of its parameters are displayed in Ref. [11].

\section{ LEPTONIC CONSTANT OF THE $B_{c}$-MESON}

\noindent \label{LC} The study of the heavy quarkonium system has played a
vital role in the development of the QCD. Some of the earliest applications
of perturbative QCD were calculations of the decay rates of charmonium [26].
These calculations were based on the assumption that, in the nonrelativistic
(NR) limit, the decay rate factors into a short-distance (SD) perturbative
part associated with the annihilation of the heavy quark and antiquark and a
long-distance (LD) part associated with the quarkonium wavefunction.
Calculations of the annihilation decay rates of heavy quarkonium have
recently been placed on a solid theoretical foundation by Bodwin {\it et al}%
. [27]. Using NRQCD [28] to seperate the SD and LD effects, Bodwin {\it et al%
}. derived a general factorization formula for the inclusive annihilation
decay rates of heavy quarkonium. The SD factors in the factorization formula
can be calculated using pQCD [19], and the LD factors are defined rigorously
in terms of the matrix elements of NRQCD that can be estimated using lattice
calculations [5]. It applies equally well to ${\rm S}$-wave, ${\rm P}$-wave,
and higher orbital-angular-momentum states, and it can be used to
incorporate relativistic corrections to the decay rates.

In the NRQCD [28] approximation for the heavy quarks, the calculation of the
leptonic decay constant for the heavy quarkonium with the two-loop accuracy
requires the matching of NRQCD currents with corresponding full-QCD
axial-vector currents [29] 
\begin{equation}
\left. {\cal J}^{\lambda }\right| _{{\rm NRQCD}}=-\chi _{b}^{\dagger }\psi
_{c}v^{\lambda }~\ \text{and \ }\left. J^{\lambda }\right| _{{\rm QCD}}=%
\overline{b}\gamma ^{\lambda }\gamma _{5}c,  \label{40}
\end{equation}
where $b$ and $c$ are the relativistic bottom and charm fields,
respectively, $\chi _{b}^{\dagger }$ and $\psi _{c}$ are the NR spinors of
anti-bottom and charm and $v^{\lambda }$ is the four-velocity of heavy
quarkonium. The \ NRQCD lagrangian describing the $B_{c}$-meson bound state
dynamics is [30]

\begin{equation}
{\cal L}_{{\rm NRQCD}}={\cal L}_{{\rm light}}+\psi _{c}^{\dagger }\left(
iD_{0}+{\bf D}^{2}/(2m_{c})\right) \psi _{c}+\chi _{b}^{\dagger }\left(
iD_{0}-{\bf D}^{2}/(2m_{b})\right) \chi _{b}+\;\cdots ,  \label{41}
\end{equation}
where ${\cal L}_{{\rm light}}$ is the relativistic lagrangian for gluons and
light quarks. The two-component spinor field $\psi _{c}$ annihilates charm
quarks, while $\chi _{b}$ creates bottom anti-quarks. The relative velocity $%
v$ of heavy quarks inside the $B_{c}$-meson provides a small parameter that
can be used as a nonperturbative expansion parameter. To express the decay
constant $f_{B_{c}}$ in terms of NRQCD matix elements we express\ $\left.
J^{\lambda }\right| _{{\rm QCD}}$ in\ terms of NRQCD fields $\psi _{c}$ and $%
\chi _{b}.$ The $\lambda =0$ current-component contributes to the matrix
element and consequently the $\left. J^{\lambda }\right| _{{\rm QCD}}$ has
the following operator expansion

\begin{equation}
\left\langle 0\right| \overline{b}\gamma ^{\lambda }\gamma _{5}c\left| B_{c}(%
{\bf P})\right\rangle =if_{B_{c}}P^{\lambda },  \label{42}
\end{equation}
where $\left| B_{c}({\bf P})\right\rangle $ is the state of the $B_{c}$%
-meson with four-momentum $P.$ Only the $\lambda =0$ component contributes
to the matrix element (42) in the rest frame of the $B_{c}$-meson. It has
the standard covariant normalization

\begin{equation}
\frac{1}{(2\pi )^{3}}\int \psi _{B_{c}}^{\ast }(p^{\prime })\psi
_{B_{c}}(p)d^{3}p=2E\delta ^{(3)}(p^{\prime }-p),  \label{43}
\end{equation}
and its phase has been chosen so that $f_{B_{c}}$ is real and positive.
Hence, the matching yields

\begin{equation}
\overline{b}\gamma ^{0}\gamma _{5}c=K_{0}\chi _{b}^{\dagger }\psi _{c}+K_{2}(%
{\bf D}\chi _{b})^{\dagger }.{\bf D}\psi _{c}+\;\cdots ,  \label{44}
\end{equation}
where $K_{0}=K_{0}(m_{c},m_{b}$) and $K_{2}=K_{2}(m_{c},m_{b}$) are Wilson
SD coefficients. They can be determined by matching perturbative
calculations of the matrix element $\left\langle 0\right| \overline{b}\gamma
^{0}\gamma _{5}c\left| B_{c}\right\rangle ,$ a contribution is mostly coming
up from the first term in 
\begin{equation}
\left. \left\langle 0\right| \overline{b}\gamma ^{0}\gamma _{5}c\left|
B_{c}\right\rangle \right| _{{\rm QCD}}=\left. \left. K_{0}\left\langle
0\right| \chi _{b}^{\dagger }\psi _{c}\left| B_{c}\right\rangle \right| _{%
{\rm NRQCD}}+K_{2}\left\langle 0\right| ({\bf D}\chi _{b})^{\dagger }.{\bf D}%
\psi _{c}\left| B_{c}\right\rangle \right| _{{\rm NRQCD}}+\;\cdots ,
\label{45}
\end{equation}
where the matrix element on the left side of (45) is taken between the
vacuum and the state $\left| B_{c}\right\rangle .$\ Hence, equation (45) can
be estimated as:\ 
\begin{equation}
\left| \left\langle 0\right| \chi _{b}^{\dagger }\psi _{c}\left|
B_{c}\right\rangle \right| ^{2}\simeq \frac{3M_{B_{c}}}{\pi }\left|
R_{1S}(0)\right| ^{2}.  \label{46}
\end{equation}
Onishchenko and Veretin [30] calculated the matrix elements on both sides of
\ equation (45) up to $\alpha _{s}^{2}$ order. In one-loop calculation, the
SD-coefficients are 
\begin{equation}
K_{0}=1\text{ \ and \ K}_{2}=-\frac{1}{8\mu ^{2}},  \label{47}
\end{equation}
with $\mu $ defined after Eq. (1). Further, Braaten and Fleming in their
work [31] calculated the perturbation correction to $K_{0}$ up to order- $%
\alpha _{s}$ (one-loop correction) as

\begin{equation}
K_{0}=1+c_{1}\frac{\alpha _{s}(\mu )}{\pi },  \label{48}
\end{equation}
with $c_{1}$ being calculated in Ref. [31] as 
\begin{equation}
c_{1}=-\left[ 2-\frac{m_{b}-m_{c}}{m_{b}+m_{c}}\ln \frac{m_{b}}{m_{c}}\right]
.  \label{49}
\end{equation}
Finally, the leptonic decay constant for the one-loop calculations is 
\begin{equation}
f_{B_{c}}^{{\rm (1-loop)}}=\left[ 1-\frac{\alpha _{s}(\mu )}{\pi }\left[ 2-%
\frac{m_{b}-m_{c}}{m_{b}+m_{c}}\ln \frac{m_{b}}{m_{c}}\right] \right]
f_{B_{c}}^{{\rm NR}},  \label{50}
\end{equation}
where the NR leptonic constant [32] is

\begin{equation}
f_{B_{c}}^{{\rm NR}}=\sqrt{\frac{3}{\pi M_{B_{c}}}}\left| R_{{\rm 1S}%
}(0)\right|  \label{51}
\end{equation}
and $\mu $ is any scale of order $m_{b}$ or $m_{c}$ of the running coupling
constant. On the other hand, the calculations of two-loop correction in the
case of vector current and equal quark masses was done in Ref. [33].
Further, Onishchenko and Veretin [30] extended the work of Ref. [33] into
the the non-equal mass case. They found an expression for the two-loop QCD
corrections to $B_{c}$-meson leptonic constant given by

\begin{equation}
K_{0}(\alpha _{s},M/\mu )=1+c_{1}(M/\mu )\frac{\alpha _{s}(M)}{\pi }%
+c_{2}(M/\mu )\left( \frac{\alpha _{s}(M)}{\pi }\right) ^{2}+\cdots ,
\label{52}
\end{equation}
where $c_{2}(M/\mu )$ is the two-loop matching coefficient and with $c_{1,2}$
are explicitly given in Eq. (49) and (cf. Ref. [30]; Eqs. (16)-(20)
therein), respectively. In the case of $B_{c}$-meson and pole quark masses $%
(m_{b}=4.8~{\rm GeV},~m_{c}=1.65~{\rm GeV}),$ they found 
\begin{equation}
f_{B_{c}}^{{\rm (2-loop)}}=\left[ 1-1.48\left( \frac{\alpha _{s}(m_{b})}{\pi 
}\right) -24.24\left( \frac{\alpha _{s}(m_{b})}{\pi }\right) ^{2}\right]
f_{B_{c}}^{{\rm NR}}.  \label{53}
\end{equation}
Therefore, the two-loop corrections are large and constitute nearly $100\%$
of one-loop correction as stated in Ref. [30].where $\mu $ is chosen as the
scale of the running coupling constant.

\section{DISCUSSION AND CONCLUSION}

\label{DAC}We use Eq. (21) to determine the position of the charmonium
center-of-gravity $\overline{M}_{\psi }({\rm 1S})$ mass spectrum. Further,
we fix the coupling constant $\alpha _{s}(m_{c})$ for each potential$.$ For
simplicity we neglect the variation of $\alpha _{s}$ with momentum in (28)
to have a common spectra for all states and scale the splitting of $%
\overline{b}c$ and $\overline{b}b$ from the charmonium value in (21). The
consideration of the variation of the effective Coulomb interaction constant
becomes especially essential for the $\Upsilon $ particle, for which $\alpha
_{s}(\Upsilon )\neq \alpha _{s}(\psi ).$\footnote{%
Kiselev {\it et al. }[34] have taken into account that $\Delta M_{\Upsilon }(%
{\rm 1S})=\frac{\alpha _{s}(\Upsilon )}{\alpha _{s}(\psi )}\Delta M_{\psi }(%
{\rm 1S})$ with $\alpha _{s}(\Upsilon )/\alpha _{s}(\psi )\simeq 3/4.$
Further, Motyka and Zalewiski [21] also found $\frac{\alpha _{s}(m_{b}^{2})}{%
\alpha _{s}(m_{c}^{2})}\simeq 11/18.$} We follow our previous works [11] to
calculate the corresponding low-lying center-of-gravity $\overline{M}%
_{\Upsilon }({\rm 1S})$ and consequently the low-lying $\overline{M}_{B_{c}}(%
{\rm 1S}).$

Table 1 reports our prediction for the Schr\"{o}dinger mass spectrum of the
four lowest $c\overline{b}$ ${\rm S}$-states together with the first three $%
{\rm P}$- and $\ {\rm D}$-states below their strong decay threshold for
different static potentials.\ Since the model is spin independent and as the
energies of some singlet states of quarkonium families have not been
measured [11,19,23], a theoretical estimates of these unknown levels
introduces uncertainty into the calculated SAD [11]. Our results in Table 1
for the $B_{c}$ and $B_{c}^{\ast }$ meson masses are in a pretty good
agreement with the other authers [1,4,7,11]. Thus, it is clear that the
Song-Lin and Turin potentials are the most preferred interactions. Here, we
report the range of the strong coupling constant at the $m_{c}$ scale we
take in our analysis $0.1985\leq \alpha _{s}(m_{c}^{2})\leq 0.320$ for all
types of potentials and $0.220\leq \alpha _{s}(m_{c}^{2})\leq 0.320$ for the
class of static potentials [11]. We point out a different choice of the
potential would in general lead to a different value of the wave function at
the origin and to a different determination of $\alpha _{s}(m_{c}^{2})$ from
the same hyperfine splitting. Further, our predictions to the $\overline{b}c$
masses of the lowest ${\rm S}$-wave (singlet and triplet) together with the
other estimations by many authors are given in Table 2. Moreover, in Table
3, we also estimate the radial wave function of the low-lying state of the $%
\overline{b}c$ system, so that

\begin{equation}
\left| R_{{\rm 1S}}(0)\right| =1.280-1.540~{\rm GeV}^{{\rm 3/2}},  \label{54}
\end{equation}
for the group of static potentials. Further, we present our calculations for
the NR leptonic constant $f_{B_{c}}^{{\rm NR}}=466_{-25}^{+19}$ ${\rm MeV}$
and $\ f_{B_{c}^{\ast }}^{{\rm NR}}=463_{-24}^{+19}$ ${\rm MeV}$ as an
estimation of the potential models without the matching [4,19]. Our results
are compared with those of other works [1,35,36]. Our calculation for the
one-loop correction:

\begin{equation}
\ f_{B_{c}}^{{\rm (1-loop)}}=408_{-14}^{+16}~{\rm MeV}\text{ and }\
f_{B_{c}^{\ast }}^{{\rm (1-loop)}}=405_{-14}^{+17}~{\rm MeV,}  \label{55}
\end{equation}
and for two-loop correction:

\begin{equation}
\ f_{B_{c}}^{{\rm (2-loop)}}=315_{-51}^{+16}~{\rm MeV}\text{ and }\
f_{B_{c}^{\ast }}^{{\rm (2-loop)}}=313_{-51}^{+26}~{\rm MeV.}  \label{56}
\end{equation}
So, our numerical value for $f_{B_{c}}^{{\rm NR}}$ agrees with the estimates
obtained in the framework of the lattice QCD result [5], $f_{B_{c}}^{{\rm NR}%
}=440\pm 20$ ${\rm MeV},$ QCD sum rules [37], potential models [1,4,19], and
from the scaling relation [34]. It indicates that the one-loop matching [33]
provides the magnitude of correction of nearly $12\%.$ Further, the most
recent calculation [29] in the heavy quark potential in the static limit of
QCD with the one-loop matching is

\begin{equation}
f_{B_{c}}^{{\rm (1-loop)}}=400\pm 15~{\rm MeV.}  \label{57}
\end{equation}
Therefore, in contrast to the discussion given in [29], we see that the
difference is not crucially large in our estimation to one-loop value in the 
$B_{c}$-meson. On the other hand, our final result of the two-loop
calculations is

\begin{equation}
f_{B_{c}}^{{\rm (2-loop)}}=315_{-50}^{+26}~{\rm MeV},  \label{58}
\end{equation}
the larger error value in (58) is due to the strongest running coupling
constant in Cornell potential. Moreover, Motyka and Zalewiski [21] also
found $f_{B_{c}}^{{\rm (1-loop)}}=435$ ${\rm MeV}$ for the ground state of $%
\overline{b}c$ quarkonium.

In the potential model, we note that slightly different additive constants
is permitted to bring up data to its center-of-gravity value. However, with
no additive constant to the Cornell potential [38], we notice that the
smaller mass values for the composing quarks of the meson leads to a rise in
the values of the potential parameters which in turn produces a notable
lower value for the leptonic constant.

Our predictions for the $\overline{b}c$ mass spectrum for the Igi-Ono
potential (type I and II) are given in Table 4$.$ Moreover, the singlet and
triplet masses together with the hyperfine splittings predicted for the two
types of this potential are also reported in Table 5. We, hereby, tested
acceptable parameters for $\Lambda _{\overline{MS}}$ from $100$ to $500~{\rm %
MeV}$ for the type I and II potentials\footnote{%
The parameters of this potential are given in Table 3 of Ref. [11].}. Small
discrepancies between our prediction and SAD experiment [11,16,19,23] can be
seen for higher states and such discrepanciesare probably seen for any
potential model and it might be related to the threshold effects or
quark-gloun mixings. The fitted set of parameters for the Igi-Ono potential
(type III) [11] are also tested in our method with $b=19$ and $(\Lambda _{%
\overline{MS}}=300~{\rm MeV}$ and also $390~{\rm MeV})$ and then\ $b=16.3$
and $\Lambda _{\overline{MS}}=300~{\rm MeV}$ which seems to be more
convenient than $\Lambda _{\overline{MS}}=500~{\rm MeV}$ used by other
authors [8]. Results of this study are also presented in Table 6. We see
that the quark masses $m_{c}$ and $m_{b}$ are sensitive to the variation of $%
\Lambda _{\overline{MS}}.$ Therefore, as $\Lambda _{\overline{MS}}$
increases the contribution of the potential (cf. e.g., Eqs. (31) and (32))
and consequently the binding energy $E_{n,l}$ term decreases which leads to
an increase in the constituent quark masses of the convenient meson, cf. Eq.
(18).

In this model, we see that the experimental $\overline{b}c$ splittings can
be reproduced for $\Lambda _{\overline{MS}}\sim 300~{\rm MeV}$ in type I, $%
\Lambda _{\overline{MS}}\sim 400$ ${\rm MeV}$ in type II (cf. Table 5) and $%
\Lambda _{\overline{MS}}\sim 300~{\rm MeV}$ in type III (cf. Table 6). We
also predicted the splittings to several ${\rm MeV}$ with the other
formalisms (c.f., Table 1 of [6]).

In Table 6, we also find that $m_{c}$ and $m_{b}$ are insensitive to the
variation of $\Lambda _{\overline{MS}}$ for this Chen-Kuang potential. This
is consistent with the conventional idea that, for heavy quarks, the
constituent quark mass is close to the current quark mass which is $\Lambda
_{\overline{MS}}$ independent. Numerical calculations show that this
potential is insensitive to $\Lambda _{\overline{MS}}$ in the range from $%
100 $ to $300~{\rm MeV}$ and as $\Lambda _{\overline{MS}}$ increases, the
potential becomes more sensetive for the ${\rm 1S}$-state only. The
theoretically calculated ${\rm n}^{{\rm 1}}{\rm S}_{{\rm 0}}$ and ${\rm n}^{%
{\rm 3}}{\rm S}_{{\rm 1}}$ hyperfine splittings for the $B_{c}$ meson in the
Chen-Kuang potential are also listed in Table 6. They are considerably
smaller than the corresponding calculated values $\Delta _{{\rm 1S}}(%
\overline{b}c)=76$ ${\rm MeV},$ and $\Delta _{{\rm 2S}}(\overline{b}c)=42$ $%
{\rm MeV}$ predicted by the quadratic formalism of Ref. [6]. Moreover,
Chen-Kuang [25] calculated $\Delta _{{\rm 1S}}(\overline{b}c)=49.9$ ${\rm MeV%
},$ and $\Delta _{{\rm 2S}}(\overline{b}c)=29.4$ ${\rm MeV}$ for their
potential with $\Lambda _{\overline{MS}}=200~{\rm MeV}$ in which the last
splitting is almost constant as $\Lambda _{\overline{MS}}$ increases. Our
theoretical calculation for $\Delta _{{\rm 1S}}(\overline{b}c)=68$ ${\rm MeV}%
,$ and $\Delta _{{\rm 2S}}(\overline{b}c)=35$ ${\rm MeV}$ for the Chen-Kuang
potential with $\Lambda _{\overline{MS}}$ runs from $100$ into $375$ ${\rm %
MeV}$. We also find $\Delta _{{\rm 1S}}(\overline{b}c)=67$ ${\rm MeV},$ and $%
\Delta _{{\rm 2S}}(\overline{b}c)=33~{\rm MeV}$ for the Igi-Ono potential
with $\Lambda _{\overline{MS}}=300~{\rm MeV}$ and$\ $ $b=16.3.$ This model
has the following features: (1) The present potential predicts smaller
calculated $\Delta _{{\rm 1S}}$ and $\Delta _{{\rm 2S}}$ than the other
potentials do for $\overline{b}\vspace{0.05cm}c$ system and our calculated $%
\Delta _{{\rm 1S}}$ and $\Delta _{{\rm 2S}}$ do not depend on $\Lambda _{%
\overline{MS}}$ more sensitively (2) The theoretical $\overline{b}c$
splitting can be repoduced for the preferred $\Lambda _{\overline{MS}}$ in
the range $100-300$ ${\rm MeV}$. Furthermore, in Table 5, for instance, we
may choose $f_{B_{c}}^{{\rm NR}}=420$ ${\rm MeV}$ with $\Lambda _{\overline{%
MS}}=300$ ${\rm MeV},$ for the type I, and $f_{B_{c}}^{{\rm NR}}=396$ ${\rm %
MeV}$ with $\Lambda _{\overline{MS}}=300$ ${\rm MeV},$ for the type II. The
result on\ $\ f_{B_{c}}$ is within the errors given by the other authors
[5,19,29]. Further, for the CK potential, we present the decay constant in
Table 6.\ 

The scaling relation (SR) for the ${\rm S}$-wave heavy quarkonia has the
form [34]

\begin{equation}
\frac{f_{n}^{2}}{M_{n}(\overline{b}c)}\left( \frac{M_{n}(\overline{b}c)}{%
M_{1}(\overline{b}c)}\right) ^{2}\left( \frac{m_{c}+m_{b}}{4\mu }\right) =%
\frac{d}{n},  \label{59}
\end{equation}
where $m_{c}$ and $m_{b}$ are the masses of heavy quarks composing the $%
B_{c} $-meson, $\mu $ is the reduced mass of quarks, and $\ d$ is a constant
independent of both the quark flavors and the level number $n.$ The value of 
$\ d$ is determined by the splitting between the ${\rm 2S}$ and ${\rm 1S}$
levels or the average kinetic energy of heavy quarks, which is independent
of the quark flavors and $n$ with the accuracy accepted. The accuracy
depends on the heavy quark masses and it is discussed in detail [34]. The
parameter value in Eq. (59), $d\simeq 55$ ${\rm MeV},$ can be extracted from
the experimentally known leptonic constants of $\psi $ and $\Upsilon .$ So,
from Table 1, the SR gives for the ${\rm 1S}$-level

\begin{equation}
\ f_{B_{c}}^{{\rm (SR)}}\simeq 444_{-23}^{+6}~{\rm MeV},  \label{60}
\end{equation}
for all static potentials used. Kiselev [29,34] estimated $\
f_{B_{c}}^{{}}=400\pm 45~{\rm MeV}$ and \ $f_{B_{c}}^{{\rm (SR)}}=385\pm 25~%
{\rm MeV}$, Narison [39] found \ $f_{B_{c}}^{{\rm (SR)}}=400\pm 25~{\rm MeV}$%
.

Overmore, we present the leptonic constants for the excited ${\rm nS}$%
-levels of the $\overline{b}c$ in Table 7. We see that our prediction $\
f_{B_{c}(2S)}^{{\rm (SR)}}=300\pm 15~{\rm MeV}$ is in good agreement with
the ones predicted by Kiselev {\it et al.} [19], $\ f_{B_{c}{\rm (2S)}}^{%
{\rm (SR)}}=280\pm 50~{\rm MeV}$ for the ${\rm 2S}$-level in the $\overline{b%
}c$ system. In Figure 1, we plot the calculated values of $B_{c}$ leptonic
constants using Eq. (59) for different potential models together with the
calculated values of the excited ${\rm nS}$-states using [34]

\begin{equation}
f_{n_{2}}=\sqrt{\frac{n_{1}}{n_{2}}}f_{n_{1}}.  \label{61}
\end{equation}
The conclusion can be drawn from the Figure 1, that the approximated values
of the excited ${\rm nS}$-states agree well with the simple scaling relation
(SR) derived from QCD sum rules for the state density. It is clear that the
estimates obtained from the potential model and SR are in good agreement to
several ${\rm MeV}$ as in Figure 1. Therefore, the difference between the
leptonic constants for the pseudoscalar and vector ${\rm 1S}$-states is
caused by the spin-dependent corrections, which are small. Numerically, we
get $\left| \ f_{B_{c}^{\ast }}^{{}}-\ f_{B_{c}}^{{}}\right| /\
f_{B_{c}^{\ast }}^{{}}<1\%.$ For the heavy quarkonia, the QCD sum rule
approximation, provides that the $\ f_{P}$ and $f_{V}$ values for the
pseudoscalar and vector states.leptonic constant is practically independent
of the total spin of quarks, so that

\begin{equation}
f_{V,n}\simeq f_{P,n}=f_{n}.  \label{62}
\end{equation}
Our numerical approximation for the decay constants of the pseudoscalar and
vector states in Tables 3, 5, and 6 is a confirmation to the last formula
(62).

In this paper, we have developed the SLET in the treatment of the $\overline{%
b}c$ system using a wide class of static and QCD-motivated potentials. For
such quarkonium potentials the method simply predicts the results of [11].
In this context, in reproducing the SAD, we used the same fitted parameters
of [11] for the sake of comparison. Further, we demonstrate to the reader
that the SLET method generates exactly the same numerical energy spectrum
for $c\overline{c},$ $b\overline{b},$and $c\overline{b}$ sresults as in the
SLNET. This refutes the claims of [13] that this method is a reformation to
SLNET and has a wider domain of applicability. Clearly, the method is simply
an alternative parallel mathematical pseudoperturbative expansion technique
having the same accuracy of [11,12,14].

\acknowledgments The author would like to thank Prof. Ramazan Sever (METU)
for his valuable discussions. He also thanks his family members for their
continuous encouragement and patience during this work.

\appendix
\bigskip

\section{SLET Parameters for the Schr\"{o}dinger Equation:}

Here, w{e list the analytic expressions of~ }$\gamma ${$^{(1)}$, }$\gamma ${$%
^{(2)}$, $\varepsilon _{i}$ and $\delta _{j}$ for the Schr\"{o}dinger
equation:}

\begin{eqnarray}
\gamma ^{(1)} &=&\left[ (1+2n_{r})\bar{\varepsilon}%
_{2}+3(1+2n_{r}+2n_{r}^{2})\bar{\varepsilon}_{4}\right] \;  \nonumber \\
&-&\omega ^{-1}\left[ \bar{\varepsilon}_{1}^{2}+6(1+2n_{r})~\bar{\varepsilon}%
_{1}\bar{\varepsilon}_{3}+(11+30n_{r}+30n_{r}^{2})\bar{\varepsilon}_{3}^{2}%
\right] ,  \label{A1}
\end{eqnarray}

\begin{eqnarray}
\gamma ^{(2)} &=&\left[ (1+2n_{r})\bar{\delta}_{2}+3(1+2n_{r}+2n_{r}^{2})%
\bar{\delta}_{4}+5(3+8n_{r}+6n_{r}^{2}+4n_{r}^{3})\bar{\delta}_{6}~\right. 
\nonumber \\
&-&\omega ^{-1}~(1+2n_{r})~\bar{\varepsilon}_{2}^{2}+12(1+2n_{r}+2n_{r}^{2})%
\bar{\varepsilon}_{2}\bar{\varepsilon}_{4}+2\bar{\varepsilon}_{1}\bar{\delta}%
_{1}  \nonumber \\
&+&2(21+59n_{r}+51n_{r}^{2}+34n_{r}^{3})\bar{\varepsilon}_{4}^{2}+6(1+2n_{r})%
\bar{\varepsilon}_{1}\bar{\delta}_{3}~  \nonumber \\
&+&30~(1+2n_{r}+2n_{r}^{2})~\bar{\varepsilon}_{1}~\bar{\delta}%
_{5}~+2~(11+30n_{r}+30n_{r}^{2})~\bar{\varepsilon}_{3}~\bar{\delta}_{3} 
\nonumber \\
&&+\left. 10(13+40n_{r}+42n_{r}^{2}+28n_{r}^{3})\bar{\varepsilon}_{3}\bar{%
\delta}_{5}+6(1+2n_{r})\bar{\varepsilon}_{3}\bar{\delta}_{1}\right] 
\nonumber \\
\text{\ }~ &+&\omega ^{-2}\left[ 4\bar{\varepsilon}_{1}^{2}\bar{\varepsilon}%
_{2}+36(1+2n_{r})\bar{\varepsilon}_{1}\bar{\varepsilon}_{2}\bar{\varepsilon}%
_{3}+8(11+30n_{r}+30n_{r}^{2})~\bar{\varepsilon}_{2}\bar{\varepsilon}%
_{3}^{2}~\right.  \nonumber \\
&+&24(1+2n_{r})\bar{\varepsilon}_{1}^{2}\bar{\varepsilon}%
_{4}+8(31+78n_{r}+78n_{r}^{2})\bar{\varepsilon}_{1}\bar{\varepsilon}_{3}\bar{%
\varepsilon}_{4}  \nonumber \\
&&+\left. 12~(57+189n_{r}+225n_{r}^{2}+150n_{r}^{3})~\bar{\varepsilon}%
_{3}^{2}~\bar{\varepsilon}_{4}\right]  \nonumber \\
&-&\omega ^{-3}\left[ 8\bar{\varepsilon}_{1}^{3}\bar{\varepsilon}%
_{3}+108(1+2n_{r})\bar{\varepsilon}_{1}^{2}\bar{\varepsilon}%
_{3}^{2}+48(11+30n_{r}+30n_{r}^{2})\bar{\varepsilon}_{1}\bar{\varepsilon}%
_{3}^{3}\right.  \nonumber \\
&+&\left. 30(31+109n_{r}+141n_{r}^{2}+94n_{r}^{3})~\bar{\varepsilon}_{3}^{4}%
\right] ,~~~~~~~~~~~~  \label{A2}
\end{eqnarray}
where 
\begin{equation}
\bar{\varepsilon _{i}}=\frac{\varepsilon _{i}}{(4\mu \omega )^{i/2}}%
,~~~~i=1,2,3,4.~~~~~  \label{A3}
\end{equation}
and

\begin{equation}
\bar{\delta _{j}}=\frac{\delta _{j}}{(4\mu \omega )^{j/2}},~~~~j=1,2,3,4,5,6.
\label{A4}
\end{equation}

\begin{equation}
\varepsilon _{1}=\frac{-(2a+1)}{2\mu }~,{~~~}\varepsilon _{2}=\frac{3(2a+1)}{%
4\mu },  \label{A5}
\end{equation}

\begin{equation}
\varepsilon _{3}=-\frac{1}{\mu }~+\frac{r_{0}^{5}V^{\prime \prime \prime
}(r_{0})}{6Q};\text{ \ \ }\varepsilon _{4}=\frac{5}{4\mu }+~\frac{%
r_{0}^{6}V^{\prime \prime \prime \prime }(r_{0})~}{24Q}\text{ }  \label{A6}
\end{equation}

\begin{equation}
\delta _{1}=-\frac{a(a+1)}{2\mu };\text{ \ }\delta _{2}=\frac{3a(a+1)}{4\mu }%
,  \label{A7}
\end{equation}

\begin{equation}
\delta _{3}=-\frac{(2a+1)}{\mu }~;{~~~}\delta _{4}=\frac{5(2a+1)}{4\mu },
\label{A8}
\end{equation}

\begin{equation}
\delta _{5}=-\frac{3}{2\mu }+\frac{r_{0}^{7}V^{\prime \prime \prime \prime
\prime }(r_{0})~}{120Q};\text{ }~\delta _{6}=\frac{7}{4\mu }+\frac{%
r_{0}^{8}V^{\prime \prime \prime \prime \prime \prime }(r_{0})~}{720Q}.
\label{A9}
\end{equation}

%\end{document}

\begin{figure}[tbp]
\caption{The $n{\rm S}$-levels leptonic constant of the $B_{c}$ system
calculated in different static potential models using SR.}
\label{autonum}
\end{figure}

\begin{table}[tbp]
\caption{The $\overline{b}c$ masses and hyperfine splittings ($\Delta _{{\rm %
nS}})$ calculated in different static potentials (in ${\rm MeV}).$}
\begin{tabular}{lllllll}
States & Refs. 1,6 & Cornell & Song-Lin & Turin & Martin & Logarithmic \\ 
\tableline$\alpha _{s}(m_{c}^{2})=$ &  & $0.320$ & $0.263$ & $0.286$ & $%
0.251 $ & $0.220$ \\ 
$m_{c}~(GeV)=$ &  & $1.840$ & $1.820$ & $1.790$ & $1.800$ & $1.500$ \\ 
$m_{b}~(GeV)=$ &  & $5.232$ & $5.199$ & $5.171$ & $5.174$ & $4.905$ \\ 
$M(\overline{b}c)$ &  &  &  &  &  &  \\ 
$1S$ & $6315$ & $6315$ & $6306$ & $6307$ & $6301$ & $6317$ \\ 
$1^{3}S_{1}$ & $6334$ & $6335$ & $6325$ & $6326$ & $6319$ & $6334$ \\ 
$1^{1}S_{0}$ & $6258$ & $6252$ & $6249$ & $6249$ & $6247$ & $6266$ \\ 
$\Delta _{1S}{}$\tablenote{$\Delta_{nS}=M(n^{3}S_{1})-M(n^{1}S_{0})$.} & $77$
& $83.5$ & $76.1$ & $76.7$ & $71.6$ & $68.0$ \\ 
$2S$ & $6873$ & $6888$ & $6875$ & $6880$ & $6892$ & $6903$ \\ 
$2^{3}S_{1}$ & $6883$ & $6897$ & $6884$ & $6889$ & $6902$ & $6911$ \\ 
$2^{1}S_{0}$ & $6841$ & $6860$ & $6850$ & $6852$ & $6865$ & $6879$ \\ 
$\Delta _{2S}$ & $42$ & $37.9$ & $34.0$ & $36.5$ & $36.7$ & $31.3$ \\ 
$3S$ & $7246$ & $7271$ & $7209$ & $7246$ & $7236$ & $7225$ \\ 
$4S$ &  & $7587$ & $7455$ & $7535$ & $7483$ & $7448$ \\ 
$1P$ & $6772$ & $6743$ & $6733$ & $6731$ & $6730$ & $6754$ \\ 
$2P$ & $7154$ & $7138$ & $7104$ & $7123$ & $7125$ & $7127$ \\ 
$3P$ &  & $7464$ & $7371$ & $7428$ & $7398$ & $7375$ \\ 
$1D$ & $7043$ & $7003$ & $6998$ & $6998$ & $7011$ & $7027$ \\ 
$2D$ & $7367$ & $7340$ & $7284$ & $7320$ & $7311$ & $7301$ \\ 
$3D$ &  & $7636$ & $7510$ & $7588$ & $7536$ & $7502$%
\end{tabular}
$\bigskip $%
\end{table}

\mediumtext

\bigskip 
\begin{table}[tbp]
\caption{The predicted $\overline{b}c$ masses of the lowest ${\rm S}$-wave
and its theoretically calculated splittings compared with the other authors
(in ${\rm MeV}).$}
\begin{tabular}{llll}
Work & $M_{B_{c}}(1^{1}S_{0})\tablenote{The experimental mass of the singlet
state is given in [2].}$ & $M_{B_{c}^{\ast }}(1^{3}S_{1})$ & $\Delta _{1S}$
\\ 
\tableline Eichten et al. [1] & $6258\pm 20$ &  &  \\ 
Colangelo and Fazio [3] & $6280$ & $6350$ &  \\ 
Baker et al. [40] & $6287$ & $6372$ &  \\ 
Roncaglia et al. [40] &  & $6320\pm 10$ &  \\ 
Godfrey et al. [1] & 6270 & 6340 &  \\ 
Bagan et al. [1,40] & 6255$\pm 20$ & $6330\pm 20$ &  \\ 
Brambilla et al. [3] &  & $6326_{-9}^{+29}$ & $60\tablenote{We remark that
we have calculated splittings from the singlet and triplet states.}$ \\ 
Baldicchi et al. [6] & $6194\sim 6292$ & $6284\sim 6357$ & $65\leq \Delta
_{1S}\leq 90$ \\ 
SLET$\tablenote{Averaging over the five values in Table 1.}$ & $%
6253_{-6}^{+13}$ & $6328_{-9}^{+7}$ & $68\leq \Delta _{1S}\leq 83$ \\ 
SLET$\tablenote{We treat Eichten and Quigg's results in the same manner
(e.g., [1]).}$ & $6258_{-11}^{+8}$ & $6333_{-14}^{+2}$ & 
\end{tabular}
\end{table}

\begin{table}[tbp]
\caption{The characteristics of the radial wave function at the origin $%
\left| R_{{\rm 1S}}(0)\right| ^{2}$ (in ${\rm GeV}^{{\rm 3}}),$ NR, one-loop
and two-loop corrections to pseudoscalar and vector decay constants of the
low-lying $B_{c}$ meson (the accuracy is 5$\%$) alculated in different
static potential models (in ${\rm MeV)}$. }
\label{table 1}
\begin{tabular}{lllllllll}
Quantity & Cornell & Song-Lin & Turin & Martin & Logarithmic & GKLT[35] & 
EFG[1] & JW[36] \\ 
\tableline$\left| \psi _{1S}(0)\right| ^{2}$ & 0.112 & 0.123 & 0.111 & 0.119
& 0.102 &  &  &  \\ 
$\left| R_{1S}(0)\right| ^{2}$ & 1.413 & 1.54 & 1.397 & 1.495 & 1.28 &  &  & 
\\ 
\tableline$f_{B_{c}}^{(NR)}$ & 464.5 & 485.1 & 462.0 & 478.0 & 441.7 & 460$%
\pm 60$ & 433 & 420$\pm 13$ \\ 
$f_{B_{c}^{\ast }}^{(NR)}$ & 461.5 & 482.2 & 459.2 & 475.3 & 439.3 & 460$\pm
60$ & 503 &  \\ 
\tableline$f_{B_{c}}^{(1-loop)}$ & 393.6\tablenote{First loop SD Wilson
coefficient for all potentials, $K_{0}=0.85-0.90$.} & 424.4 & 399.6 & 421.2
& 399.3 &  &  &  \\ 
$f_{B_{c}}^{(2-loop)}$ & 264.1\tablenote{Second loop SD Wilson coefficient
for all potentials, $K_{0}=0.57-0.77$.} & 333.0 & 296.6 & 339.1 & 340.9 &  & 
&  \\ 
$f_{B_{c}^{\ast }}^{(1-loop)}$ & 391.0 & 421.9 & 397.1 & 418.8 & 397.2 &  & 
&  \\ 
$f_{B_{c}^{\ast }}^{(2-loop)}$ & 262.3 & 331.0 & 294.8 & 337.2 & 339.0 &  & 
& 
\end{tabular}
\end{table}

\mediumtext

\bigskip

\begin{table}[tbp]
\caption{The $\overline{b}c$ mass spectra predicted for various $\Lambda _{%
\overline{MS}}$ using Igi-Ono (type I and II) potential (in ${\rm MeV}).$}
\begin{tabular}{lllllll}
&  &  &  & $\Lambda _{\overline{MS}}$ &  &  \\ 
States & [6,24] & $100$ & $200$ & $300$ & $400$ & $500$ \\ 
\tableline$b=20$\tablenotemark[1] & $\alpha _{s}=$ & $0.1985$ & 0.217 & 0.238
& 0.250 & 0.262 \\ 
$1S$ & $6327$ & $6329$ & $6318$ & $6310$ & $6316$ & $6327$ \\ 
$2S$ & $6906$ & $6915$ & $6904$ & $6881$ & $6880$ & $6901$ \\ 
$3S$ & $7246$ & $7264$ & $7242$ & $7244$ & $7241$ & $7252$ \\ 
$4S$ &  & 7508 & 7522 & 7545 & 7542 & 7552 \\ 
$1P$ & $6754$ & $6755$ & $6744$ & $6733$ & $6732$ & $6742$ \\ 
$2P$ & $7154$ & $7144$ & $7131$ & $7125$ & $7122$ & $7134$ \\ 
$1D$ & $7028$ & $7029$ & $7017$ & $7004$ & $7000$ & $7010$ \\ 
$2D$ & $7367$ & $7334$ & $7327$ & $7327$ & $7323$ & $7333$ \\ 
\tableline$b=5$\tablenotemark[2] & $\alpha _{s}=$ & 0.1985 & 0.227 & 0.230 & 
0.2405 &  \\ 
$1S$ & $6327$ & $6331$ & $6324$ & $6316$ & $6307$ &  \\ 
$2S$ & $6906$ & $6914$ & $6898$ & $6910$\tablenotemark[3] & $6918$ &  \\ 
$3S$ & $7246$ & $7258$ & $7277$ & $7236$ & $7201$\tablenotemark[3] &  \\ 
$4S$ &  & 7521 & 7517 & 7478 & 7500 &  \\ 
$1P$ & 6754 & 6756 & 6743 & 6737 & 6730 &  \\ 
$2P$ & $7154$ & $7142$ & $7138$ & $7134$ & $7120$ &  \\ 
$1D$ & $7028$ & $7029$ & $7015$ & $7012$\tablenotemark[3] & $7007$ &  \\ 
$2D$ & $7367$ & $7335$ & $7323$\tablenotemark[3] & $7314$ & $7316$ & 
\end{tabular}
\tablenotetext[1]{$c_{0}=-0.022$ to $-0.031$ $MeV$.}%
\tablenotetext[2]{$c_{0}=-0.019$ to $-0.026$ $MeV$.}%
\tablenotetext[3]{Carried
out to the second correction order.}
\end{table}

\bigskip 
\begin{table}[tbp]
\caption{The $\overline{b}c$ mass spectrum, splittings and leptonic constant
predicted for various $\Lambda _{\overline{MS}}$ using Igi-Ono (type I and
II) potential (in ${\rm MeV}).$}
\begin{tabular}{llllll}
&  &  & $\Lambda _{\overline{MS}}$ &  &  \\ 
States & 100 & 200 & 300 & 400 & 500 \\ 
\tableline Type I &  &  &  &  &  \\ 
$1^{3}S_{1}$ & 6343 & 6334 & 6327 & 6334 & 6344 \\ 
$1^{1}S_{0}$ & 6287 & 6272 & 6259 & 6263 & 6274 \\ 
$\Delta _{1S}$ & 56.3 & 62.0 & 68.3 & 71.1 & 69.8 \\ 
$\left| R_{1S}(0)\right| ^{2}$ & 0.826 & 1.005 & 1.156 & 1.19 & 1.114 \\ 
$f_{B_{c}}^{NR}$ & 354.1 & 391.1 & 420.0 & 426.0 & 411.7 \\ 
$f_{B_{c}}^{(1-loop)}$ & 328.1 & 356.5 & 376.8 & 379.2 & 364.4 \\ 
$f_{B_{c}}^{(2-loop)}$ & 290.0 & 306.1 & 311.7 & 306.5 & 287.1 \\ 
$f_{B_{c}^{\ast }}^{NR}$ & 352.6 & 389.2 & 417.7 & 423.6 & 409.4 \\ 
$f_{B_{c}^{\ast }}^{(1-loop)}$ & 326.7 & 354.8 & 374.7 & 377.1 & 362.4 \\ 
$f_{B_{c}^{\ast }}^{(2-loop)}$ & 288.7 & 304.6 & 310.0 & 304.7 & 285.6 \\ 
\tableline Type II &  &  &  &  &  \\ 
$1^{3}S_{1}$ & 6345 & 6340 & 6331 & 6323 &  \\ 
$1^{1}S_{0}$ & 6288 & 6279 & 6269 & 6259 &  \\ 
$\Delta _{1S}$ & 56.7 & 60.6 & 61.8 & 64.4 &  \\ 
$\left| R_{1S}(0)\right| ^{2}$ & 0.819 & 0.891 & 1.03 & 1.204 &  \\ 
$f_{B_{c}}^{NR}$ & 352.7 & 368.2 & 396.0 & 428.6 &  \\ 
$f_{B_{c}}^{(1-loop)}$ & 327.1 & 334.9 & 357.4 & 382.1 &  \\ 
$f_{B_{c}}^{(2-loop)}$ & 289.1 & 283.0 & 300.1 & 314.4 &  \\ 
$f_{B_{c}^{\ast }}^{NR}$ & 351.2 & 366.4 & 394.1 & 426.4 &  \\ 
$f_{B_{c}^{\ast }}^{(1-loop)}$ & 325.6 & 333.3 & 355.7 & 380.1 &  \\ 
$f_{B_{c}^{\ast }}^{(2-loop)}$ & 287.8 & 281.7 & 298.6 & 312.8 & 
\end{tabular}
\end{table}

\bigskip 
\begin{table}[tbp]
\caption{The $\overline{b}c$ mass spectrum, splittings and leptonic constant
predicted for various $\Lambda _{\overline{MS}}$ using Igi-Ono (type III)
and Chen-Kuang potentials (in ${\rm MeV}).$}
\begin{tabular}{lllllll}
State &  & IO (III) &  &  & CK &  \\ 
\tableline$b=$ & $16.3$ & 19 & 19 & 5.1 & 5.1 & 5.1 \\ 
$\Lambda _{\overline{MS}}=$ & $300$ & $300$ & $390$ & $100-300$ & 350 & 375
\\ 
$\alpha _{s}=$ & $0.250$ & $0.2505$ & $0.2205$ & $0.270$ & $0.270$ & $0.270$
\\ 
$1S$ & $6309$ & $6309$ & 6297 & $6324$ & $6372$ & 6354 \\ 
$2S$ & $6880$ & $6870$ & 6877 & $6880$ & $6880$ & 6880 \\ 
$3S$ & $7247$ & $7236$ & $7254$ & $7258$ & $6258$ & 6258 \\ 
$4S$ & $7553$ & $7541$ & $7563$ & $7570$ & $7570$ & 7570 \\ 
$1P$ & $6725$ & $6721$ & $6737$ & $6723$ & $6723$ & 6723 \\ 
$2P$ & $7124$ & $7114$ & $7135$ & $7127$ & $7127$ & 7127 \\ 
$3P$ & 7441 & 7429 & 7452 & 7452 & 7452 & 7452 \\ 
$1D$ & $6997$ & 6990 & $7013$ & $6993$ & 6993 & 6993 \\ 
$2D$ & $7328$ & 7317 & $7341$ & $7332$ & 7332 & 7332 \\ 
$3D$ & 7613 & 7599 & 7624 & 7625 & 7625 &  \\ 
$1^{3}S_{1}$ & $6326$ & $6327$ & $6315$ & $6341$ & 6389 & 6371 \\ 
$1^{1}S_{0}$ & $6259$ & $6258$ & $6243$ & $6273$ & 6321 & 6304 \\ 
$\Delta _{1S}$ & $67.3$ & $68.6$ & $72.6$ & $67.8$ & 67.8 & 67.7 \\ 
$\left| R_{1S}(0)\right| ^{2}$ & 1.115 & 1.119 & 1.339 & 1.017 & $1.017$ & 
1.017 \\ 
$f_{B_{c}}^{NR}$ & $412.4$ & $413.2$ & $452.6$ & $393.5$ & $392.0$ & 392.5
\\ 
$f_{B_{c}}^{(1-loop)}$ & $367.1$ & $367.2$ & $408.0$ & $347.5$ & $346.2$ & 
346.6 \\ 
$f_{B_{c}}^{(2-loop)}$ & $296.4$ & $294.0$ & $345.4$ & $269.0$ & $268.0$ & 
268.4 \\ 
$f_{B_{c}^{\ast }}^{NR}$ & 410.2 & 411.0 & 450.0 & $391.4$ & $390.0$ & 390.5
\\ 
$f_{B_{c}^{\ast }}^{(1-loop)}$ & 365.2 & 365.2 & 405.7 & 345.6 & $344.3$ & 
344.8 \\ 
$f_{B_{c}^{\ast }}^{(2-loop)}$ & 294.8 & 292.4 & 343.4 & 267.6 & $266.6$ & 
266.9
\end{tabular}
\end{table}

\mediumtext

\bigskip 
\begin{table}[tbp]
\caption{The $n{\rm S}$-levels leptonic constant of the $\overline{b}c$
system$,$ calculated in different static potential models (the accuracy is $%
3-7\%$), in ${\rm MeV}$, using the SR. }
\label{table7}
\begin{tabular}{llllll}
Quantity & Cornell & Song-Lin & Turin & Martin & Logarithmic \\ 
\tableline$f_{1S}$ & 449.6 & 450.4 & 448.0 & 448.8 & 420.9 \\ 
$f_{2S}$ & 305.8 & 305.0 & 303.3 & 303.5 & 284.7 \\ 
$f_{3S}$ & 243.0 & 243.2 & 241.3 & 241.8 & 227.2 \\ 
$f_{4S}$ & 206.0 & 207.1 & 204.9 & 205.9 & 193.8
\end{tabular}
\end{table}

\bigskip \bigskip

\bigskip

\end{document}